\documentclass[%
reprint,
superscriptaddress,
amsmath,amssymb,
aps,
]{revtex4-2}

\usepackage{graphicx}
\usepackage{dcolumn}
\usepackage{bm}

\begin{document}


\title{First Detection of Photons with Energy Beyond 100 TeV \\ from an Astrophysical Source}

\author{M.~Amenomori}
\affiliation{Department of Physics, Hirosaki University, Hirosaki 036-8561, Japan }
\author{Y.~W.~Bao}
\affiliation{School of Astronomy and Space Science, Nanjing University, Nanjing 210093, China }
\author{X.~J.~Bi}
\affiliation{Key Laboratory of Particle Astrophysics, Institute of High Energy Physics, Chinese Academy of Sciences, Beijing 100049, China }
\author{D.~Chen}
\affiliation{National Astronomical Observatories, Chinese Academy of Sciences, Beijing 100012, China }
\author{T.~L.~Chen}
\affiliation{Physics Department of Science School, Tibet University, Lhasa 850000, China }
\author{W.~Y.~Chen}
\affiliation{Key Laboratory of Particle Astrophysics, Institute of High Energy Physics, Chinese Academy of Sciences, Beijing 100049, China }
\author{Xu~Chen$^{\dagger}$}
\affiliation{Key Laboratory of Particle Astrophysics, Institute of High Energy Physics, Chinese Academy of Sciences, Beijing 100049, China }
\affiliation{University of Chinese Academy of Sciences, Beijing 100049, China }
\author{Y.~Chen}
\affiliation{School of Astronomy and Space Science, Nanjing University, Nanjing 210093, China }
\author{Cirennima}
\affiliation{Physics Department of Science School, Tibet University, Lhasa 850000, China }
\author{S.~W.~Cui}
\affiliation{Department of Physics, Hebei Normal University, Shijiazhuang 050016, China }
\author{Danzengluobu}
\affiliation{Physics Department of Science School, Tibet University, Lhasa 850000, China }
\author{L.~K.~Ding}
\affiliation{Key Laboratory of Particle Astrophysics, Institute of High Energy Physics, Chinese Academy of Sciences, Beijing 100049, China }
\author{J.~H.~Fang}
\affiliation{Key Laboratory of Particle Astrophysics, Institute of High Energy Physics, Chinese Academy of Sciences, Beijing 100049, China }
\affiliation{University of Chinese Academy of Sciences, Beijing 100049, China }
\author{K.~Fang}
\affiliation{Key Laboratory of Particle Astrophysics, Institute of High Energy Physics, Chinese Academy of Sciences, Beijing 100049, China }
\author{C.~F.~Feng}
\affiliation{Department of Physics, Shandong University, Jinan 250100, China }
\author{Zhaoyang~Feng}
\affiliation{Key Laboratory of Particle Astrophysics, Institute of High Energy Physics, Chinese Academy of Sciences, Beijing 100049, China }
\author{Z.~Y.~Feng}
\affiliation{Institute of Modern Physics, SouthWest Jiaotong University, Chengdu 610031, China }
\author{Qi~Gao}
\affiliation{Physics Department of Science School, Tibet University, Lhasa 850000, China }
\author{Q.~B.~Gou}
\affiliation{Key Laboratory of Particle Astrophysics, Institute of High Energy Physics, Chinese Academy of Sciences, Beijing 100049, China }
\author{Y.~Q.~Guo}
\affiliation{Key Laboratory of Particle Astrophysics, Institute of High Energy Physics, Chinese Academy of Sciences, Beijing 100049, China }
\author{H.~H.~He}
\affiliation{Key Laboratory of Particle Astrophysics, Institute of High Energy Physics, Chinese Academy of Sciences, Beijing 100049, China }
\author{Z.~T.~He}
\affiliation{Department of Physics, Hebei Normal University, Shijiazhuang 050016, China }
\author{K.~Hibino}
\affiliation{Faculty of Engineering, Kanagawa University, Yokohama 221-8686, Japan }
\author{N.~Hotta}
\affiliation{Faculty of Education, Utsunomiya University, Utsunomiya 321-8505, Japan }
\author{Haibing~Hu}
\affiliation{Physics Department of Science School, Tibet University, Lhasa 850000, China }
\author{H.~B.~Hu}
\affiliation{Key Laboratory of Particle Astrophysics, Institute of High Energy Physics, Chinese Academy of Sciences, Beijing 100049, China }
\author{J.~Huang$^{\S}$}
\affiliation{Key Laboratory of Particle Astrophysics, Institute of High Energy Physics, Chinese Academy of Sciences, Beijing 100049, China }
\author{H.~Y.~Jia}
\affiliation{Institute of Modern Physics, SouthWest Jiaotong University, Chengdu 610031, China }
\author{L.~Jiang}
\affiliation{Key Laboratory of Particle Astrophysics, Institute of High Energy Physics, Chinese Academy of Sciences, Beijing 100049, China }
\author{H.~B.~Jin}
\affiliation{National Astronomical Observatories, Chinese Academy of Sciences, Beijing 100012, China }
\author{F.~Kajino}
\affiliation{Department of Physics, Konan University, Kobe 658-8501, Japan }
\author{K.~Kasahara}
\affiliation{Research Institute for Science and Engineering, Waseda University, Tokyo 169-8555, Japan }
\author{Y.~Katayose}
\affiliation{Faculty of Engineering, Yokohama National University, Yokohama 240-8501, Japan }
\author{C.~Kato}
\affiliation{Department of Physics, Shinshu University, Matsumoto 390-8621, Japan }
\author{S.~Kato}
\affiliation{Institute for Cosmic Ray Research, University of Tokyo, Kashiwa 277-8582, Japan }
\author{K.~Kawata$^{*}$}
\affiliation{Institute for Cosmic Ray Research, University of Tokyo, Kashiwa 277-8582, Japan }
\author{M.~Kozai}
\affiliation{Institute of Space and Astronautical Science, Japan Aerospace Exploration Agency (ISAS/JAXA), Sagamihara 252-5210, Japan}
\author{Labaciren}
\affiliation{Physics Department of Science School, Tibet University, Lhasa 850000, China }
\author{G.~M.~Le}
\affiliation{National Center for Space Weather, China Meteorological Administration, Beijing 100081, China }
\author{A.~F.~Li}
\affiliation{School of Information Science and Engineering, Shandong Agriculture University, Taian 271018, China }
\affiliation{Department of Physics, Shandong University, Jinan 250100, China }
\affiliation{Key Laboratory of Particle Astrophysics, Institute of High Energy Physics, Chinese Academy of Sciences, Beijing 100049, China }
\author{H.~J.~Li}
\affiliation{Physics Department of Science School, Tibet University, Lhasa 850000, China }
\author{W.~J.~Li}
\affiliation{Key Laboratory of Particle Astrophysics, Institute of High Energy Physics, Chinese Academy of Sciences, Beijing 100049, China }
\affiliation{Institute of Modern Physics, SouthWest Jiaotong University, Chengdu 610031, China }
\author{Y.~H.~Lin}
\affiliation{Key Laboratory of Particle Astrophysics, Institute of High Energy Physics, Chinese Academy of Sciences, Beijing 100049, China }
\affiliation{University of Chinese Academy of Sciences, Beijing 100049, China }
\author{B.~Liu}
\affiliation{School of Astronomy and Space Science, Nanjing University, Nanjing 210093, China }
\author{C.~Liu}
\affiliation{Key Laboratory of Particle Astrophysics, Institute of High Energy Physics, Chinese Academy of Sciences, Beijing 100049, China }
\author{J.~S.~Liu}
\affiliation{Key Laboratory of Particle Astrophysics, Institute of High Energy Physics, Chinese Academy of Sciences, Beijing 100049, China }
\author{M.~Y.~Liu}
\affiliation{Physics Department of Science School, Tibet University, Lhasa 850000, China }
\author{Y.-Q.~Lou}
\affiliation{Physics Department, Astronomy Department and Tsinghua Center for Astrophysics, Tsinghua-National Astronomical Observatories of China joint Research Center for Astrophysics, Tsinghua University, Beijing 100084, China}
\author{H.~Lu}
\affiliation{Key Laboratory of Particle Astrophysics, Institute of High Energy Physics, Chinese Academy of Sciences, Beijing 100049, China }
\author{X.~R.~Meng}
\affiliation{Physics Department of Science School, Tibet University, Lhasa 850000, China }
\author{H.~Mitsui}
\affiliation{Faculty of Engineering, Yokohama National University, Yokohama 240-8501, Japan }
\author{K.~Munakata}
\affiliation{Department of Physics, Shinshu University, Matsumoto 390-8621, Japan }
\author{Y.~Nakamura}
\affiliation{Key Laboratory of Particle Astrophysics, Institute of High Energy Physics, Chinese Academy of Sciences, Beijing 100049, China }
\author{H.~Nanjo}
\affiliation{Department of Physics, Hirosaki University, Hirosaki 036-8561, Japan }
\author{M.~Nishizawa}
\affiliation{National Institute of Informatics, Tokyo 101-8430, Japan }
\author{M.~Ohnishi}
\affiliation{Institute for Cosmic Ray Research, University of Tokyo, Kashiwa 277-8582, Japan }
\author{I.~Ohta}
\affiliation{Sakushin Gakuin University, Utsunomiya 321-3295, Japan }
\author{S.~Ozawa}
\affiliation{Research Institute for Science and Engineering, Waseda University, Tokyo 169-8555, Japan }
\author{X.~L.~Qian}
\affiliation{Department of Mechanical and Electrical Engineering, Shandong Management University, Jinan 250357, China }
\author{X.~B.~Qu}
\affiliation{College of Science, China University of Petroleum, Qingdao, 266555, China }
\author{T.~Saito}
\affiliation{Tokyo Metropolitan College of Industrial Technology, Tokyo 116-8523, Japan }
\author{M.~Sakata}
\affiliation{Department of Physics, Konan University, Kobe 658-8501, Japan }
\author{T.~K.~Sako}
\affiliation{Institute for Cosmic Ray Research, University of Tokyo, Kashiwa 277-8582, Japan }
\author{Y.~Sengoku}
\affiliation{Faculty of Engineering, Yokohama National University, Yokohama 240-8501, Japan }
\author{J.~Shao}
\affiliation{Key Laboratory of Particle Astrophysics, Institute of High Energy Physics, Chinese Academy of Sciences, Beijing 100049, China }
\affiliation{Department of Physics, Shandong University, Jinan 250100, China }
\author{M.~Shibata}
\affiliation{Faculty of Engineering, Yokohama National University, Yokohama 240-8501, Japan }
\author{A.~Shiomi}
\affiliation{College of Industrial Technology, Nihon University, Narashino 275-8576, Japan }
\author{H.~Sugimoto}
\affiliation{Shonan Institute of Technology, Fujisawa 251-8511, Japan }
\author{M.~Takita$^{\ddagger}$}
\affiliation{Institute for Cosmic Ray Research, University of Tokyo, Kashiwa 277-8582, Japan }
\author{Y.~H.~Tan}
\affiliation{Key Laboratory of Particle Astrophysics, Institute of High Energy Physics, Chinese Academy of Sciences, Beijing 100049, China }
\author{N.~Tateyama}
\affiliation{Faculty of Engineering, Kanagawa University, Yokohama 221-8686, Japan }
\author{S.~Torii}
\affiliation{Research Institute for Science and Engineering, Waseda University, Tokyo 169-8555, Japan }
\author{H.~Tsuchiya}
\affiliation{Japan Atomic Energy Agency, Tokai-mura 319-1195, Japan }
\author{S.~Udo}
\affiliation{Faculty of Engineering, Kanagawa University, Yokohama 221-8686, Japan }
\author{H.~Wang}
\affiliation{Key Laboratory of Particle Astrophysics, Institute of High Energy Physics, Chinese Academy of Sciences, Beijing 100049, China }
\author{H.~R.~Wu}
\affiliation{Key Laboratory of Particle Astrophysics, Institute of High Energy Physics, Chinese Academy of Sciences, Beijing 100049, China }
\author{L.~Xue}
\affiliation{Department of Physics, Shandong University, Jinan 250100, China }
\author{K.~Yagisawa}
\affiliation{Faculty of Engineering, Yokohama National University, Yokohama 240-8501, Japan }
\author{Y.~Yamamoto}
\affiliation{Department of Physics, Konan University, Kobe 658-8501, Japan }
\author{Z.~Yang}
\affiliation{Key Laboratory of Particle Astrophysics, Institute of High Energy Physics, Chinese Academy of Sciences, Beijing 100049, China }
\author{A.~F.~Yuan}
\affiliation{Physics Department of Science School, Tibet University, Lhasa 850000, China }
\author{L.~M.~Zhai}
\affiliation{National Astronomical Observatories, Chinese Academy of Sciences, Beijing 100012, China }
\author{H.~M.~Zhang}
\affiliation{Key Laboratory of Particle Astrophysics, Institute of High Energy Physics, Chinese Academy of Sciences, Beijing 100049, China }
\author{J.~L.~Zhang}
\affiliation{Key Laboratory of Particle Astrophysics, Institute of High Energy Physics, Chinese Academy of Sciences, Beijing 100049, China }
\author{X.~Zhang}
\affiliation{School of Astronomy and Space Science, Nanjing University, Nanjing 210093, China }
\author{X.~Y.~Zhang}
\affiliation{Department of Physics, Shandong University, Jinan 250100, China }
\author{Y.~Zhang}
\affiliation{Key Laboratory of Particle Astrophysics, Institute of High Energy Physics, Chinese Academy of Sciences, Beijing 100049, China }
\author{Yi~Zhang}
\affiliation{Key Laboratory of Particle Astrophysics, Institute of High Energy Physics, Chinese Academy of Sciences, Beijing 100049, China }
\author{Ying~Zhang}
\affiliation{Key Laboratory of Particle Astrophysics, Institute of High Energy Physics, Chinese Academy of Sciences, Beijing 100049, China }
\author{Zhaxisangzhu}
\affiliation{Physics Department of Science School, Tibet University, Lhasa 850000, China }
\author{X.~X.~Zhou}
\affiliation{Institute of Modern Physics, SouthWest Jiaotong University, Chengdu 610031, China }
\collaboration{The Tibet AS$\gamma$ Collaboration}


\date{April 4, 2019; Submitted to the Physical Review Letters}

\begin{abstract}
We report on the highest energy photons from the Crab Nebula observed
by the Tibet air shower array with the underground
water-Cherenkov-type muon detector array.  Based on the criterion of
muon number measured in an air shower, we successfully suppress
99.92\% of the cosmic-ray background events with energies $E>100$~TeV.
As a result, we observed 24 photon-like events with $E>100$~TeV
against 5.5 background events, which corresponds to 5.6$\sigma$
statistical significance.  This is the first detection of photons with
$E>100$~TeV from an astrophysical source.
\end{abstract}

\maketitle


\section{\label{Intro} Introduction}

The Crab Nebula powered by a central spinning pulsar and
a fluctuating magnetized relativistic pulsar wind \cite{Lou93}
is one of the most energetic astrophysical sources in the entire
sky, and the energy spectrum of photons from it
has been measured in a wide energy range from radio up to
nearly 100~TeV. However, the photons of energy $E>100$~TeV have never
been detected so far.

Photons in the TeV energies from the Crab Nebula have been observed with many ground-based gamma-ray experiments
\cite{Weekes89,Konopelko96,Tanimori98,Amenomori99,Aharonian06,Albert08,Abdo12,Bartoli13,Aliu14,Abeysekara17}.
Among them, the HEGRA experiment had obtained the energy spectrum 
which can be approximately fitted by a single power-law shape $E^{-p}$ in
the highest energy range up to 75~TeV \cite{Aharonian04}.  
Alternatively, the H.E.S.S. experiment characterized
the observed energy spectrum by $E^{-p} {\rm exp}(-E/E_{\rm c})$,
with the index $p=2.39\pm0.03$ and the exponential cutoff energy $E_{\rm c} = (14.3\pm2.1)$~TeV
between 440~GeV and 40~TeV \cite{Aharonian06}.  
At sub-PeV energies between 141~TeV and 646~TeV,
the CASA-MIA experiment had set stringent flux upper limits by an
air shower (AS) array and the underground muon detector (MD) array
\cite{Borione97}.

The Tibet AS$\gamma$ experiment achieved the first successful
observation of the Crab Nebula in the multi-TeV region in 1999, using
the Tibet AS array with an area of 5,175~m$^2$
\cite{Amenomori99}.  Subsequently, the Tibet-III array with an area of
22,050~m$^2$ has been operating since 1999.  With
this array, we measured the energy spectrum of
the Crab Nebula at energies between 1.7 and 40~TeV \cite{Amenomori09}.  In 2007, a
prototype underground water-Cherenkov-type MD with a
detection area of 100~m$^2$ for two cells was built beneath the Tibet-III
array.  By this MD, we determined the most stringent flux upper
limit for Crab photons $>140$~TeV \cite{Amenomori15}.

Since the beginning of 2014, a water-Cherenkov-type MD array with
a total area of $\sim$3,400~m$^2$ started operation.
In this Letter, we report on the photon spectrum of the Crab in
the highest energy range of 3 to $\sim$400~TeV observed by
the Tibet AS array with this new MD array. 

\section{Experiment of AS and MD Arrays}

\begin{figure}[ht]
\includegraphics[width=8.5cm]{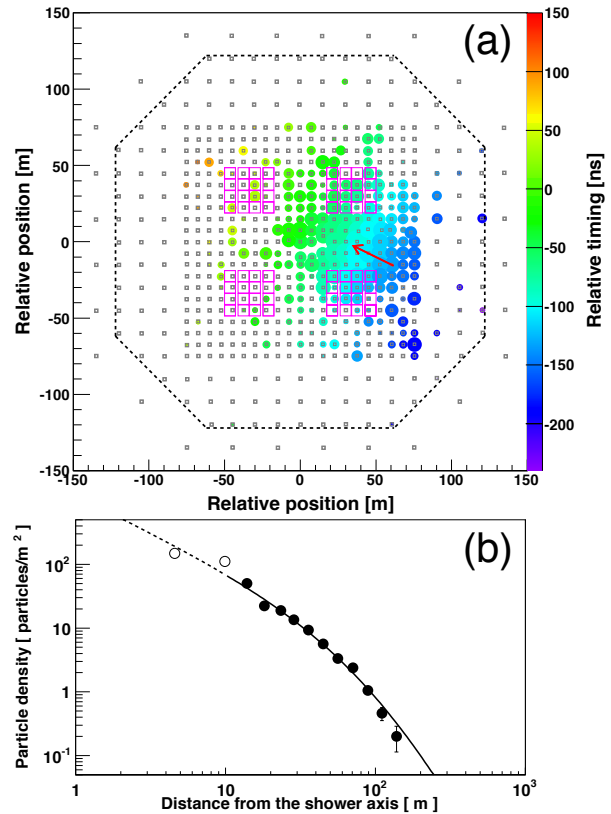}
\caption{(a) An event display of the observed photon-like AS 
  of energy 251~TeV. The size and color of each circle represent the logarithmic
  particle density and the relative timing in each detector, respectively. 
  The arrow head and direction indicate the AS core location and incident direction, respectively.
  Dots and open squares denote scintillation detectors and underground
  MDs, respectively.  The enclosed
  area by the dashed line indicates the fiducial area of the AS array.  (b)
  Lateral distribution of the photon-like shower event in panel (a).
  The solid circles and curve show the experimental data and fitting result by the Nishimura-Kamata-Greisen (NKG)
  function to the data recorded by detectors more than 10~m apart from the AS axis.  The dashed curve and open circles
  are an extrapolation of the NKG function fitting and the unused data within 10~m from the AS axis.
\label{fig_1}}
\end{figure}

The Tibet AS array has been continuously observing very-high-energy cosmic
rays above TeV at Yangbajing (90.522$^{\circ}$E, 30.102$^{\circ}$N;
Altitude 4300~m) in Tibet, China \cite{Amenomori09}.  The current
AS array, covering an
area of 65,700~m$^{2}$, consists of 597 plastic scintillation detectors indicated by small dots in Fig.~\ref{fig_1}, each
with 0.5~m$^{2}$ detection area.  This array detects the
electromagnetic components in an AS, such as $e^{\pm}$'s and $\gamma$'s,
and each detector measures the arrival times and densities of the detected particles.  
With these data, the arrival direction and energy of the primary cosmic ray
are then reconstructed event by event.

The Tibet MD array consists of 64 water-Cherenkov-type detectors
located at 2.4~m underground beneath the AS array as shown by open
squares in Fig.~\ref{fig_1}(a).  Each detector is a waterproof concrete
cell filled with water of 1.5~m in depth, 7.35~m $\times$ 7.35~m in
area, viewed by a 20-inch-diameter downward-facing photomultiplier
tube (PMT) on the ceiling.  The inner walls and floor are covered by white Tyvek
sheets to efficiently collect water Cherenkov light produced by muons in water.
The electromagnetic component is
shielded by the soil overburden corresponding to $\sim$19 radiation
lengths, while the energy threshold for muons to penetrate the soil is approximately 1~GeV.  A
primary photon induced AS produces much less muons than a
primary cosmic ray induced AS does \cite{Gaisser91}.  The Tibet MD array thus enables us to efficiently
discriminate a cosmic-ray background event from a photon signal by means of
counting muon number in an AS.

Operating these two arrays in parallel, we collected AS and muon data during 719
live days from 2014 February to 2017 May.

\section{Air Shower Data Analysis}

The arrival direction of an AS is reconstructed using the
relative timing recorded at each scintillation detector.  
The color and size of a circle in Fig.~\ref{fig_1} (a) represent the
relative timing ($\tau$) and the number of particle density ($\rho$) measured by each detector in a sample AS event, respectively.  First, we
obtain AS core location weighted by $\rho$.
The $\tau$'s in the AS front are fitted by a conical shape, and its
cone angle is optimized by the Monte Carlo (MC) simulations depending on the AS size.  
The arrow head and direction indicate
the reconstructed core position and incident direction of the AS, respectively.  
The angular resolutions (50\% containment) are estimated to be approximately 0.5$^{\circ}$ and
0.2$^{\circ}$ for 10~TeV and 100~TeV photons, respectively.

The secondary particles in an AS 
deposit energy proportional to $\rho$, in a scintillator.  At each detector, $\rho$ is 
obtained from the PMT output charge divided by the single particle peak \cite{Amenomori08}
which is monitored every 20 minutes to correct temperature dependence of each
detector gain.
For $E>10$~TeV, the energy of each AS is reconstructed using the lateral
distribution of $\rho$ shown in Fig.~\ref{fig_1} (b) by an example.
As an energy estimator, we use $S50$
defined as $\rho$ at a perpendicular distance of 50~m from
the AS axis in the best-fit NKG
function \cite{Kawata17}.  The conversion from $S50$ to the energy is optimized as a function of zenith angle by the MC simulation.
The energy resolutions with $S50$, which depend on AS core
location and zenith angle (see Fig.~S$1$ in the supplemental material), are roughly estimated to be 40\% at 10~TeV and 20\%
at 100~TeV. 
At $E<10$~TeV, we estimate energy directly from $\Sigma\rho$,
which is the sum of particle density measured by each scintillation detector,
because it is difficult to apply the NKG fitting 
due to a limited number of hit detectors. The energy resolution with $\Sigma\rho$
is estimated to be $\sim$100\% at 3~TeV.
The absolute energy scale uncertainty was
estimated to be 12\% from the westward shift of the Moon's shadow center
caused by the geomagnetic field \cite{Amenomori09}.

Muons and a part of hadronic components in an AS penetrate into the underground MD array, 
while the electromagnetic cascade rapidly attenuates in the soil above.
The number of muons detected in an MD ($N_{\rm \mu}$) is obtained from the output charge divided by 
the single muon peak which is monitored every 20 minutes.  The sum of detected particles in
all 64 MDs ({\it i.e.} $\Sigma N_{\rm \mu}$) is taken as the parameter to distinguish 
photons from cosmic rays that generate ASs.

\begin{figure}[t]
\includegraphics[width=8.5cm]{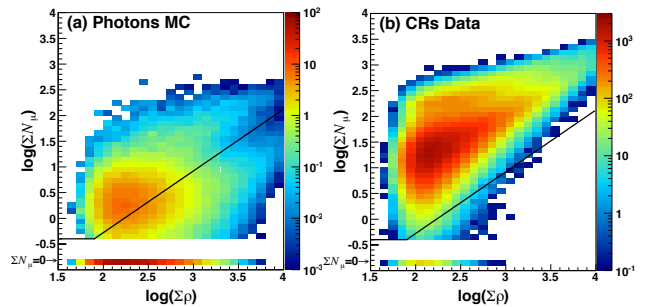}
\caption{ Distribution of the number of muons ($\Sigma N_{\rm \mu}$) measured
  by the MD array as a function of the sum of particle density ($\Sigma\rho$) measured by the AS
  array for (a) photon signals generated by the MC simulation, (b)
  cosmic-ray events extracted from the real data. The color and the solid
  lines represent the number of events and the optimized muon-cut condition. $\Sigma N_{\rm \mu} = 0$
  is plotted at log$(\Sigma N_{\rm \mu}) = -0.8$ on the vertical axis.
\label{fig_2}}
\end{figure}

The trigger condition of an AS is issued at any
4-fold coincidence of scintillation detectors within the area enclosed by the dashed lines in
Fig.~\ref{fig_1}, each recording more than 0.6 particles.  
The AS event selections and energy estimation below 10~TeV are 
carried out in the same way as our previous works \cite{Amenomori09} 
except for the muon cut. 
At $E>10$~TeV, the following event selection criteria are imposed
to ensure better energy resolution: 
(1) zenith angle of arrival direction ($\theta$) is
$<40^{\circ}$; (2) number of available detectors for the AS
reconstruction is $\geq16$; (3) among 6 detectors
recording the largest $\rho$ values, 5 are contained in
the fiducial area enclosed by the dashed lines in Fig.~\ref{fig_1};
(4) log($S50$) is $>-1.2$; (5) age parameter ($s$) in
the best-fit NKG function is between 0.3 and 1.3. (6) 
$\Sigma N_{\rm \mu} < 2.1 \times 10^{-3} (\Sigma\rho)^{1.2}$ or $\Sigma N_{\rm \mu} < 0.4$ 
as indicated by solid lines in Fig.~\ref{fig_2}.
This muon cut condition is optimized 
by the MC simulations for the observation of the photon induced ASs
(see next section).

In order to estimate the background contribution from cosmic rays, we
adopt the Equi-Zenith Angle method which was used in our previous
works \cite{Amenomori03,Amenomori09}.  The number of cosmic-ray background events is
estimated from the number of events averaged over 20 off-source 
windows located at the same zenith angle as the on-source window (but at different azimuth angle).
The radius of the on/off-source window $R_{\rm sw}$ is set to
$R_{\rm sw}(\Sigma\rho)=6.9/\sqrt{\Sigma\rho}$ ($^{\circ}$) \cite{Amenomori03}. In order to efficiently extract signals
in the higher energy region at low background level, the lower limit
of $R_{\rm sw}$ is set to 0.5$^{\circ}$, corresponding to
$\sim$90\% containment of photons with $E>100$~TeV.

\section{MC Simulations}

We simulate AS events in the atmosphere, 
using the CORSIKA code v7.4000
\cite{Corsika98} with EPOS-LHC \cite{Pierog15} for the high-energy hadronic
interaction model and FLUKA code v2011.2b \cite{Ferrari05,Bohflen14} for the low-energy hadronic interaction model.
The differential power-law index of photon spectrum takes to be $-3.0$ above 0.3~TeV.  The AS
cores are located randomly within 300~m from the AS array
center.  The generated secondary particles in an AS are fed
into the detector simulation of the AS array developed by using the GEANT4 code v4.10.00
\cite{Agostinelli03}.  The energy deposit and timing at each
scintillation detector are converted to measurable charge
and timing values considering the
detector response and the calibrations.  The simulated dataset is
analyzed in the same way as the experimental data to reconstruct the
energy and arrival direction of the primary cosmic rays that initiate ASs.
We verified that our MC simulations reproduce the
experimental AS data very well, regarding the cosmic-ray $\Sigma\rho$
spectrum and zenith angle distribution as well as the air shower
reconstruction (angular resolution, pointing accuracy, and absolute
energy scale) determined from the Moon's shadow analysis  \cite{Amenomori03,Amenomori09}.

The basic idea and the detailed study on the MC simulation of
the underground MD array are described in \cite{Sako09}.
The electromagnetic and hadronic cascades in the overburden soil,
as well as the ray tracing of Cherenkov lights emitted in the water cell are
simulated by the GEANT4 code. The number of photoelectrons detected in all MDs
is converted to $\Sigma N_{\rm \mu}$ in the same way as the experimental data.

The muon cut condition described in the preceding section is optimized using the MC events of primary
photons and real cosmic-ray data extracted from the same declination band
as the Crab, but outside the Crab region, because
$\Sigma N_{\rm \mu}$ distribution induced by the cosmic ray might depend on
the hadronic interaction and chemical composition models assumed in the MC simulation.
Figure~\ref{fig_2} (a) and (b) show $\Sigma N_{\rm \mu}$ as a function of
$\Sigma\rho$ for photon signals generated by the MC simulation, and the real
cosmic-ray background events, respectively.  The absolute photon flux of the
Crab is assumed in the photon MC simulation.  The muon cut indicated by
the solid lines in Fig.~\ref{fig_2} is determined to maximize the
figure of merit $N_{\rm \gamma}^{\rm MC}/\sqrt{N_{\rm \gamma}^{\rm
    MC}+N_{\rm CR}^{\rm DATA}}$, where $N_{\rm \gamma}^{\rm MC}$ and
$N_{\rm CR}^{\rm DATA}$ denote the expected number of photons by the
  MC simulation and the number of background events in the data after the muon cut.
The events remaining after the muon cut are regarded as the photon-like events.
It is noted that the electron/photon discrimination is very difficult
so far, because both electron and photon induced air showers
consist of similar electromagnetic component. However, the expected
ratio of electrons to cosmic rays at 10 TeV is estimated to be less
than 0.1\%, because of a steeper power-law index of the electron spectrum above 1 TeV.
Furthermore, the diffuse electrons should be
subtracted as isotropic background events by the Equi-Zenith Angle
method.

\section{Results and Discussion}

\begin{figure}[t]
\includegraphics[width=8.5cm]{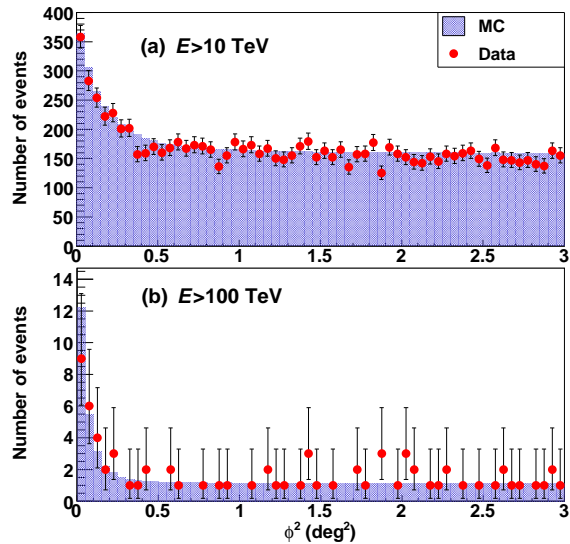}
\caption{ Distribution of events as a function of the square of the
  incident angle ($\phi^{2}$) measured from the Crab direction.
  The filled circles and the solid histograms stand for
  the experimental data and the MC with (a) $E>10$~TeV and (b) $E>100$~TeV,
  respectively.  \label{fig_3}}
\end{figure}

The cosmic-ray background events are reduced to 1.1\%
above 10~TeV with 70\% of the photons remaining after the muon cut. We
detected a clear excess from the Crab with 18.3$\sigma$ statistical
significance.  Figure~\ref{fig_3} shows a $\phi^{2}$ distribution of
events, where $\phi$ denotes incident angle 
measured from the Crab direction for (a) $E>10$~TeV and
(b) $E>100$~TeV, respectively.  The simulated distributions
assuming a point source (blue shaded histograms) well reproduce the
experimental data (red solid circles).
The 2D photon images are available in Fig.~S$2$ in the supplemental material.

With $E>100$~TeV, the cosmic-ray 
background events are significantly excluded by the muon cut down to
0.08\% with 90\% of the photons remaining.  As a result,
we detected 24 photon-like events with $E>100$~TeV against
5.5 cosmic-ray background events (the excess is 18.5 events), corresponding to 5.6$\sigma$ statistical
significance, where the $\alpha$ parameter used for calculating the Li-Ma significance is 0.05 (see Eq.~$(17)$ of \cite{Li83}).

The contamination from the lower energies below 100~TeV
due to the finite energy resolution is estimated to be 3 events.
The observed muon distribution for $E>100$~TeV after the muon cut 
is consistent with that estimated by the photon MC simulation as shown in Fig.~S$3$ in the supplemental material.

\begin{figure}[t]
\includegraphics[width=8.5cm]{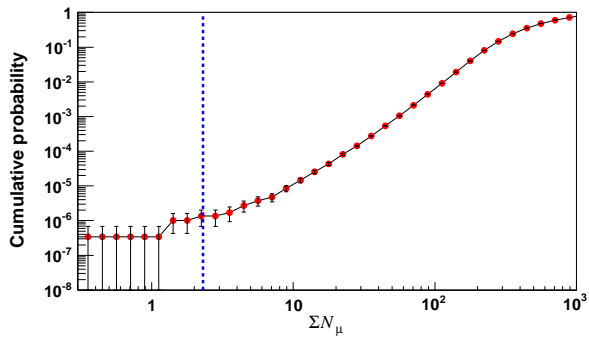}
\caption{ Cumulative probability ($P_{\rm \mu}$) of $\Sigma N_{\rm \mu}$ for
  cosmic-ray events above 251~TeV, which are recorded under the
  similar geometries ($\theta \pm 5^{\circ}$ and $r_{\rm core}
    \pm 30~{\rm m}$) as the 251~TeV photon-like event in Fig.~\ref{fig_1}.  The
  vertical dashed line indicates $\Sigma N_{\rm \mu} = 2.3$ detected in the
  251~TeV photon-like event.
\label{fig_4}}
\end{figure}

Figure~\ref{fig_1} (a) and
(b) show the event display and the lateral distribution,
respectively, of a typical photon-like event observed with $E = 251$~TeV. 
The total uncertainty of the energy, $\Delta E$, is defined as the quadratic sum of
the absolute energy-scale error (12\%) \cite{Amenomori09} and the
energy resolution which is estimated by using events of MC simulations with the same $\theta$
and core distance from the AS array center ($r_{\rm core}$), and 
is estimated to be $^{+45}_{-43}$~TeV for this 251~TeV photon-like event.
The MD array recorded only $\Sigma N_{\rm \mu} = 2.3$ for this event, while the average $\Sigma N_{\rm \mu}$ of
the cosmic-ray background events with similar energy is approximately 500.
In order to evaluate the probability of such a low $\Sigma N_{\rm \mu}$ event, we
examine the $\Sigma N_{\rm \mu}$ distribution in the real cosmic-ray events.
Figure~\ref{fig_4} shows the cumulative probability of $\Sigma N_{\rm \mu}$
with cosmic-ray events above 251~TeV,
which are recorded with the similar $\theta$ and $r_{\rm core}$ as the 251~TeV photon-like event.
We estimate the
chance probability of cosmic-ray event with $\Sigma N_{\rm \mu}<2.3$ to be
$P_{\rm \mu} = 1.36\times10^{-6}$, and the number of background events from
the Crab without the muon cut is $N^{\rm NoCut}_{\rm BG} =
1224$ events above 251~TeV. Therefore, the probability $P_{\rm CR}$ of
misidentifying a cosmic-ray event from the Crab as the observed
251~TeV photon-like event is calculated to be $P_{\rm CR} = P_{\rm \mu} \times N^{\rm NoCut}_{\rm
  BG} / {\rm 1~event} = 1.7\times10^{-3}$.  Above 250~TeV, we
found 4 photon-like events against 0.8 cosmic-ray background events 
corresponding to 2.4$\sigma$ statistical significance.  
The contamination from the lower energies below 250~TeV
due to the finite energy resolution is estimated to be 0.4 events.
The $P_{\rm CR}$'s and other parameters of these four events are summarized in
Table~\ref{tab1}. The $P_{\rm CR}$'s in Table 1 indicate that 3 events 
are highly photon-like, while the highest energy event is a borderline 
photon-like event which is consistent with 
cosmic-ray background event with a probability of 0.23.

\begin{table}[h]
\caption{\label{tab1} 
Probability of misidentifying cosmic-ray events from the Crab as a photon-like event ($P_{\rm CR}$) for each of four photon-like
events above 250~TeV together with other reconstructed values. $\theta$ and $r_{\rm core}$ are the zenith angle and core distance from the AS array center, respectively.}
\begin{ruledtabular}
\begin{tabular}{cccccccc}
\textrm{$E$}&\textrm{$\Delta E$}&\textrm{$\Sigma\rho$}&\textrm{$\Sigma N_{\rm \mu}$}&\textrm{$\theta$}&\textrm{$r_{\rm core}$}&\textrm{$\phi^{2}$}&\textrm{$P_{\rm CR}(>E)$}\\
\textrm{(TeV)}&\textrm{(TeV)}&\textrm{}&\textrm{}&\textrm{($^{\circ}$)}&\textrm{(m)}&\textrm{(deg$^{2}$)}&\textrm{}\\
\colrule
251 & $^{+46}_{-43}$ & 3248 & 2.3 & 29.8 & 35.1 & 0.00 & $1.7\times10^{-3}$\\
313 & $^{+58}_{-54}$ & 2440 & 5.5 & 27.5 & 94.6 & 0.03 & $2.2\times10^{-2}$\\
449 & $^{+112}_{-97}$ & 2307 & 11.3 & 35.4 & 93.3 & 0.12 & $2.9\times10^{-2}$\\
458 & $^{+83}_{-78}$ & 2211 & 21.5 & 27.5 & 111.6 & 0.18 & 0.23\\
\end{tabular}
\end{ruledtabular}
\end{table}

\begin{figure}[t]
\includegraphics[width=8.5cm]{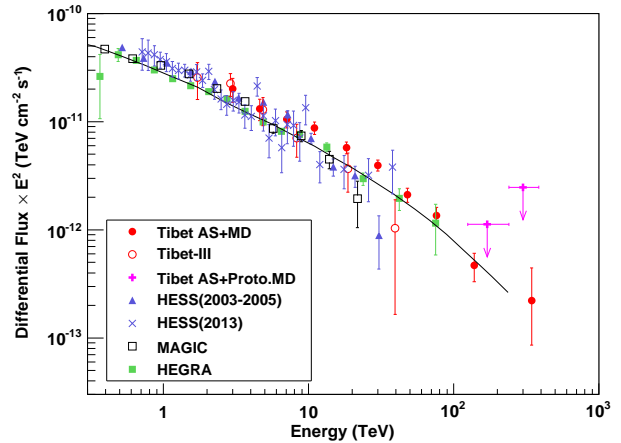}
\caption{ Differential energy spectrum of Crab photons.  The red solid circles
  and open circles show fluxes observed with the Tibet AS+MD and
  Tibet-III array \cite{Amenomori09}, respectively.  The magenta arrows with plus mark show the previous upper limits
  at the 90\% confidence level given by the Tibet AS and the 100~m$^2$ prototype MD \cite{Amenomori15}. The green squares, open squares, blue solid
  triangles and blue crosses show flux points observed with three
  Cherenkov telescopes : HEGRA \cite{Aharonian04}, MAGIC \cite{Aleksic15}, HESS
  \cite{Aharonian06} and HESS during the gamma-ray flare periods
  detected by ${\it Fermi}$-LAT in 2013 \cite{Abramowski14}, respectively.  The solid curve
  is a model fit for the IC scattering of various seed photons
  \cite{Aharonian04}.
\label{fig_5}}
\end{figure}

Finally, Fig.~\ref{fig_5} shows the differential energy spectrum of
Crab photons.  The red solid circles indicate the energy spectrum
measured by the Tibet AS+MD array, featuring by a single
power law of $(dN/dE) = (1.49\pm0.09)\times10^{-15} (E/{\rm40~TeV})^{
  -2.91\pm0.04}$ ${\rm cm}^{-2} \ {\rm s}^{-1} \ {\rm TeV}^{-1}$ in
the energy range between 3 and $\sim$400~TeV.  
The unfolding procedure of the spectrum is basically the same as
employed in the previous works \cite{Amenomori09}, while the energy
resolutions are improved at higher energies. The energy bin purities
evaluated from the smearing by the energy resolution is estimated to be 83\%
(86\%) for $100<E\le250$~TeV ($250<E\le630$~TeV). In each bin, spillover
fraction from lower and higher energy bins are 14\% (12\%) and 3\%
(2\%), respectively, for $100<E\le250$~TeV ($250<E\le630$~TeV).
We found no clear
evidence for the exponential cutoff below 100~TeV.  The
spectrum measured by the Tibet-III array up to 40~TeV \cite{Amenomori09}
is shown by the red open circles. Both
spectra are mutually consistent with each other in the overlapping energy range within statistical errors.  The
H.E.S.S. experiment measured spectra in 2003-2005 \cite{Aharonian06}
and 2013 during the gamma-ray flare periods detected by ${\it Fermi}$-LAT
\cite{Abramowski14}.  The former appears to favor an
exponential cutoff shape, while the latter seems to extend 
the power-law trend.  The spectrum measured by the Tibet AS+MD well
follows the data of the HEGRA experiment, and extends
to the sub-PeV energy regime without cutoff sign.
The integral fluxes observed by the Tibet AS+MD array are also calculated to be
$F{\rm (>100 TeV)} = (3.29^{+1.06}_{-0.87}) \times 10^{-15} \ {\rm
  cm}^{-2} {\rm s}^{-1}$ and $F{\rm (>250TeV)} =
(5.72^{+5.72}_{-3.48}) \times 10^{-16} \ {\rm cm}^{-2} {\rm s}^{-1}$,
respectively, which are consistent with and lower than the previous upper
limits given by the CASA-MIA experiment \cite{Borione97} and Tibet AS
with 100~m$^{2}$ prototype MD \cite{Amenomori15}, respectively.

The emission mechanism for the multi-TeV photons from the Crab
is thought to be the inverse-Compton (IC) scattering of ambient seed
photons by relativistic electrons \cite{Atoyan96}.  The 
model energy spectra have been calculated based on the IC scattering of
various seed photons, such as the synchrotron emission, far-infrared,
cosmic microwave background (CMB) radiation, and so on.  The solid
curve in Fig.~\ref{fig_5} is the best-fit model for the
HEGRA data \cite{Aharonian04}.  
Our spectrum above 100 TeV is consistent with the simple extrapolation
up to a few hundred TeV of the HEGRA spectrum by the IC model.
The origin of the photons exceeding 100~TeV might be 
sub-PeV electrons scattering off the low-energy CMB photons in the Crab Nebula.
Such ultra-high energy electrons and positrons can
be produced and sustained by forward and reverse
magnetohydrodynamic shocks in inhomogeneous, magnetically
striped, relativistic Crab pulsar wind mainly composed of
electrons and positrons by number \cite{Lou96,Lou98}.

\section{Conclusions}

We successfully observed 24 photon-like events with $E>100$~TeV
from the Crab Nebula against 5.5 cosmic-ray background events, corresponding to 5.6$\sigma$
statistical significance, with the Tibet AS array and
the underground water-Cherenkov-type MD array.  This is the first
detection of the highest energy photons beyond 100~TeV from an astrophysical source,
and thus opens up the sub-PeV window in astronomy.

\begin{acknowledgments}
The collaborative experiment of the Tibet Air Shower Arrays has been
conducted under the auspices of the Ministry of Science and Technology
of China and the Ministry of Foreign Affairs of Japan.  This work was
supported in part by a Grant-in-Aid for Scientific Research on
Priority Areas from the Ministry of Education, Culture, Sports,
Science, and Technology, by Grants-in-Aid for Science Research from
the Japan Society for the Promotion of Science in Japan.  This work is
supported by the National Key R\&D Program of China
(No.2016YFE0125500).  This work is supported by the Grants from the
National Natural Science Foundation of China
(Nos.11533007,11673041,11873065 and 11773019), and by the Key
Laboratory of Particle Astrophysics, Institute of High Energy Physics,
CAS. This work is supported by the Chinese Ministry of Education.
This work is supported by the joint research program of the Institute
for Cosmic Ray Research (ICRR), the University of Tokyo.
\end{acknowledgments}

\noindent
Corresponding authors:\\
$^{*}$ kawata@icrr.u-tokyo.ac.jp\\
$^{\dagger}$ chenxu@ihep.ac.cn\\
$^{\ddagger}$ takita@icrr.u-tokyo.ac.jp\\
$^{\S}$ huangjing@ihep.ac.cn




%

\end{document}